\documentclass[twocolumn,10pt,showkeys,showpacs]{revtex4} 
\usepackage{amsmath}
\usepackage{bm}
\usepackage{graphicx}
\PassOptionsToPackage{sort&compress}{natbib}

\begin{document}

\title{Induced Bremsstrahlung  by light in graphene}
\author{Cristian Villavicencio$^1$ and Alfredo Raya$^2$}


\affiliation{$^1$Departamento de Ciencias B\'asicas, Facultad de Ciencias,
Universidad del B\'io-B\'io, Casilla 447, Chill\'an, Chile.\\
Instituto de F\'isica y Matem\'aticas, Universidad Michoacana de San 
Nicol\'as de Hidalgo, Edificio C-3, Ciudad Universitaria, C.P. 58040, Morelia, 
Michoac\'an, Mexico.}

\begin{abstract}
\noindent
{\bf Abstract:}
We study the generation of an electromagnetic current in monolayer graphene immersed in a weak perpendicular magnetic field and radiated with linearly polarized monochromatic light. Such a current emits Bremsstrahlung radiation with the same amplitude above and below the plane of the sample, in the latter case consistent with the small amount of light absorption in the material. This mechanism could be an important contribution for the reflexion of light phenomenon in graphene.

\noindent
{\bf Resumen:}
Estudiamos la generaci\'on de una corriente electromagn\'etica en una monocapa de grafeno inmersa en un campo magn\'etico d\'ebil perpendicular y radiada con luz monocrom\'atica. Esta corriente emite radiaci\'on de Bremsstrahlung con la misma amplitud por arriba y abajo del plano de la muestra, en el \'ultimo caso consistente con la peque\~na cantidad de absorci\'on de luz en el material. Este mecanismo puede ser una contribuci\'on importante para el fen\'omeno de reflexi\'on de luz en grafeno.

\end{abstract}
\keywords{Graphene,  Bremsstrahlung radiation, Induced electromagnetic currents.
\\Descriptores: Grafeno, Radiaci\'on de  Bremsstrahlung, Corrientes electromagn\'eticas inducidas. }
\pacs{73.22.Pr,74.25.N-,78.67.Wj}
\maketitle

\begin{figure}
\centering
\includegraphics[scale=1]{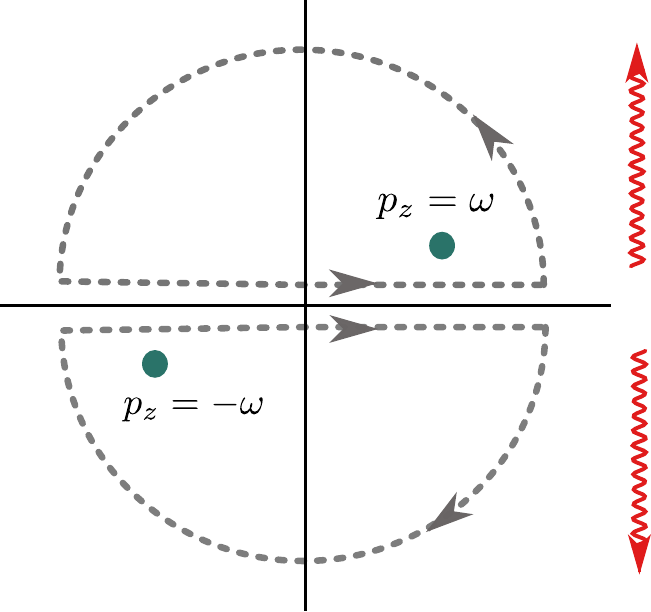}
\caption{Displaced poles for $A_\mu^{\rm rad}$ and the contour of integrations chosen to enclose each pole.
The chosen contour, above or below the real axis in the complex $p_z$ plane, produces a traveling wave propagating above or below the graphene membrane.}
\label{fig1}
\end{figure}

Graphene~\cite{novoselovnat2005,kimnat2005} continues to provide an excellent laboratory to explore fundamental physics, let alone its technological applications~(see, for instance, Ref.~\cite{GeimReview}). The gapless pseudo-relativistic Dirac nature of it  charge carriers at low energies is responsible for many of the outstanding properties of this 2 dimensional (2D) crystal that has given rise to a new era of materials science~\cite{graphenebook}. 
High electric and thermal conductivity, stiffness and flexibility of graphene flakes are a few examples in this connection. On top of these properties, the transparency of its membranes is remarkable: Only 2.3$\%$ of visible light is absorbed by a single membrane~\cite{GeimTransparency}. This rate has been verified under a number of experimental conditions~\cite{GrasseeFaradayRot}. 
On the other hand, many theoretical approaches have been used in the past to explain the rate of light absorption in graphene including quantum field theoretical methods~\cite{Fialkovs2012qft, Fialkovs2009, Fialkovs2012Faraday, VHLR1, VHRS,FLMR}. Yet, the underlying mechanism  that explains  light absorption is less transparent. In this communication, we explore the possibility for such a mechanism to be explained in terms of Bremsstrahlung radiation. The issue of Bremsstrahlung in graphene has recently been addressed~\cite{brehem1,brehem2,brehem3}. 
Here we consider the situation where a graphene membrane is located in the $z=0$ plane and is subjected to a weak magnetic field of strength $B$ oriented perpendicularly to the sample, and then is radiated with 
an  electromagnetic plane-wave, monochromatic (frequency $\omega$) and linearly-polarized, traveling from the top ($z>0$). 
The vector potential describing this plane-wave is
\begin{equation}
A_\mu(r)= -g_{\mu 2}\frac{ E_0}{i\omega}e^{i(kz+\omega t)} \;,
\end{equation}
such that the electric field $\bm{E}$ (intensity $E_0$) is oriented in the $y$ direction. We use the metric tensor $g=\textrm{diag}(1,-1,-1,-1)$ and units where $\hbar = c = 1$ and, therefore, $k=\omega$ in vacuum. 
In this form,
\begin{equation}
\begin{array}{cc}
\bm{E}  = E_0 \,\bm{\hat y}\,e^{i\omega (z+t)},
&\qquad
\bm{B}  = E_0 \,\bm{\hat x}\,e^{i\omega( z+t)}.
\end{array}
\end{equation}
The electric current generated in the graphene sample induced by the external electromagnetic wave is defined through the polarization tensor as
\begin{equation}
j_\mu(r) = \int d^4r'\; \Pi_{\mu\nu}(r-r')A_\mathrm{pl}^{\nu}(r')\;.
\end{equation}
Here, $A_\mathrm{pl}(r)= A^{*}(r)\delta(z)$ is the conjugated vector potential constrained to the plane, and  
$\Pi$ is the polarization tensor, also constrained to the membrane. For weak external magnetic fields, it is defined in momentum space as~\cite{VHLR1}
\begin{eqnarray}
&& \hspace{-1cm}\Pi_{\mu\nu}(p)  =
i\frac{\sqrt{\tilde p^2}}{v_F^2}\frac{\alpha\pi}{2}\eta_{\mu a}\Bigg\{
\nonumber \\ &&  \hspace{-.5cm}
\left[1+\left(\frac{eB}{\tilde p^2}\right)^2\left(1-5\frac{p_0^2}{\tilde p^2}\right)\right]
\left(g^{ab}-\frac{\tilde p^a\tilde p^b}{\tilde p^2}\right)
\nonumber\\ && \hspace{-.5cm}
+2\left(\frac{eB}{\tilde p^2}\right)^2\left(1-\frac{p_0^2}{\tilde p^2}\right)
\left(g^{ab}_\perp-\frac{\tilde p^a_\perp\tilde p^b_\perp}{\tilde p^2_\perp}\right)
\Bigg\}\eta_{b\nu}\;,
\nonumber\\
\end{eqnarray}
where $\tilde p^a = \eta^{a\mu}p_\mu$ and $g^{ab}=\eta^{a\mu}g_{\mu\nu}\eta^{\nu b}$, being the projector matrix $\eta_{a\mu}$ is
\begin{equation}
[\eta]_{b\nu} =
\left(\begin{array}{cccc}
1   &   0   &   0   &   0   \\
0   &   v_F &   0   &   0   \\
0   &   0   &   v_F &   0
\end{array}\right).
\end{equation}
Notice that $\eta$ slightly differs from the definition in~\cite{Fialkovs2012qft,Fialkovs2009,Fialkovs2012Faraday,VHLR1,VHRS}. 
Here, it also acts as a projector in Lorenz indexes into the graphene plane.
In our conventions, along the graphene membrane we use $g_\perp =\textrm{diag}(0,-1,-1)$ and $\tilde p_\perp = (0,v_Fp_x,v_Fp_y)$. With these definitions, it is straightforward to find that
\begin{equation}
j_\mu(r) = -g_{\mu 2}\, E\,e^{-i\omega t}\delta(z)\;,
\end{equation}
where
\begin{equation}
E=E_0\frac{\alpha\pi}{2}\left[1-4\left(\frac{eB}{\omega^2}\right)^2\right]\;.
\end{equation}
From the above results, following Ref.~\cite{peskin}, the radiated electromagnetic field can be obtained by solving Maxwell's equations in Lorentz gauge $\partial^2A_\mu^\mathrm{rad}(r) = j_\mu(r)$, which in Fourier space correspond to $-p^2A_\mu^\mathrm{rad}(p)= j_\mu(p)$.
The radiated electromagnetic vector potential $A^{\rm rad}$ is then the inverse inverse Fourier transformation of  $-j_\mu(p)/p^2$.
The integral involved to obtain it is ill-defined, as it contains poles on the real axis. In accordance to the prescription of retarded boundary conditions, we consider displacing the poles slightly from the real axis in the form shown in Fig.~\ref{fig1}.
After the integration in the $p_0$, $p_x$ and $p_y$ components, the radiation vector potential is 
\begin{equation}
A_\mu^\mathrm{rad}(r) 
=-g_{\mu2}\, E \int\frac{dp_z}{2\pi}\frac{e^{i(p_z z-\omega t)}}{p_z^2-\omega^2-i\epsilon}\;.
\end{equation}
\noindent
Furthermore, selecting the contour of integration as shown in Fig.~\ref{fig1}, integrations over momentum components are readily done. We thus  can identify the emitted radiation above ($+$) and and below ($-$) the plane as
\begin{equation}
A_\mu^\mathrm{rad\ (\pm)}(r)=
g_{\mu2}\,\frac{E}{2i\omega}e^{i\omega(\pm  z-t )}\;.
\end{equation}
The corresponding emitted electromagnetic fields are, therefore,
\begin{eqnarray}
\bm{E}_+  &=& (E/2) \, \bm{\hat y}\, e^{i\omega (z-t)}\\
\bm{B}_+  &=& (E/2) \, \bm{\hat x}\, e^{i\omega( z-t)+i\pi},\\
\bm{E}_-  &=& (E/2) \, \bm{\hat y}\, e^{-i\omega (z+t)},\\
\bm{B}_-  &=& (E/2) \, \bm{\hat x}\,e^{-i\omega(z+t)}.
\end{eqnarray}
We can straightforwardly see that the emitted wave below and above the plane carry half of the re-emitted intensity.
Thus, the total radiated energy density is ${\cal E} = E^2/2$.
Although the electric component of the wave emission in the upper side of the sample is the same in the lower side, the magnetic component of the emitted wave acquires a phase shift of $\pi$. 
The radiated emission diminishes as the external magnetic field strength increases, reflecting the tendency of the magnetic field to deflect charged-particles trajectories.

The experimental measurements of perpendicularly  incident light on graphene~\cite{GeimTransparency,GrasseeFaradayRot} as well as the theoretical description~\cite{Fialkovs2012qft, Fialkovs2009, Fialkovs2012Faraday, VHLR1, VHRS} coincide in that the order of the opacity rate is  $\sim\alpha$. The experimental arrangement is basically radiation from the top, detection from the bottom, and comparison between them (transparency). The theoretical description, in turn, accounts for the absorption rate. But none of this assumptions make explicit reference to the reflection rate. In this work we describe the re-emission of light by Bremsstrahlung, half transmitted and half reflected. The reflected amount is of order $\alpha^2$, much lower than the experimental detection of light opacity. However, there is not much done in order to detect perpendicular reflection and theoretical estimations~\cite{hermosa} reproduce null reflection when the incident beam is perpendicular to the graphene layer. Bremsstrahlung at the moment could be the main mechanism responsible for perpendicular light reflection.


\section*{Acknowledgements}
This work was supported by FONDECYT (Chile) under grant numbers 1150847, 1130056, 1150471 and , 1170107, CIC-UMSNH (M\'exico)  grant number 4.22, and CONACyT (M\'exico) grant number 256494.  
We acknowledge the group {\em F\'isica de Altas Energ\'ias} at UBB.



\begin{thebibliography}{10}
\expandafter\ifx\csname url\endcsname\relax
  \def\url#1{\texttt{#1}}\fi
\expandafter\ifx\csname urlprefix\endcsname\relax\def\urlprefix{URL }\fi

\bibitem{novoselovnat2005}
K.~Novoselov, A.~K. Geim, S.~Morozov, D.~Jiang, M.~Katsnelson, I.~Grigorieva,
  S.~Dubonos, A.~Firsov, Two-dimensional gas of massless dirac fermions in
  graphene, Nature 438~(7065) (2005) 197--200.

\bibitem{kimnat2005}
Y.~Zhang, Y.-W. Tan, S.~HL, K.~P, Experiental observation of the quantum hall
  effect and berry's phase in graphene, Nature 438~(7065) (2005) 201--204.

\bibitem{GeimReview}
A.~K. Geim, K.~S. Novoselov, The rise of graphene, Nature materials 6~(3)
  (2007) 183--191.

\bibitem{graphenebook}
M.~I. Katsnelson, M.~I. Katsnelson, Graphene: carbon in two dimensions,
  Cambridge University Press, 2012.

\bibitem{GeimTransparency}
R.~Nair, P.~Blake, A.~Grigorenko, K.~Novoselov, T.~Booth, T.~Stauber, N.~Peres,
  A.~Geim, Fine structure constant defines visual transparency of graphene,
  Science 320~(5881) (2008) 1308--1308.

\bibitem{GrasseeFaradayRot}
I.~Grassee, J.~Levallois, A.~L. Walter, M.~Ostler, A.~Bostwick, E.~Rotenberg,
  T.~Seyller, D.~Van Der~Marel, A.~B. Kuzmenko, Giant faraday rotation in
  single-and multilayer graphene, Nature Physics 7~(1) (2011) 48--51.

\bibitem{Fialkovs2012qft}
I.~Fialkovsky, D.~Vassilevich, Quantum field theory in graphene, International
  Journal of Modern Physics A 27~(15) (2012) 1260007.

\bibitem{Fialkovs2009}
I.~Fialkovsky, D.~Vassilevich, Parity-odd effects and polarization rotation in
  graphene, Journal of Physics A: Mathematical and Theoretical 42~(44) (2009)
  442001.

\bibitem{Fialkovs2012Faraday}
I.~Fialkovsky, D.~Vassilevich, Faraday rotation in graphene, The European
  Physical Journal B 85~(11) (2012) 1--10.

\bibitem{VHLR1}
D.~Valenzuela, S.~Hern{\'a}ndez-Ortiz, M.~Loewe, A.~Raya, Graphene transparency
  in weak magnetic fields, Journal of Physics A: Mathematical and Theoretical
  48~(6) (2015) 065402.

\bibitem{VHRS}
S.~Hern{\'a}ndez-Ortiz, D.~Valenzuela, A.~Raya, S.~S{\'a}nchez-Madrigal, Light
  absorption in deformed graphene, International Journal of Modern Physics B
  30~(1650084) (2015) 1--9.
	
\bibitem{FLMR} H.~Falomir, M.~Loewe, E. Mu\~noz and A.~Raya, Optical Transparency in an effective model of Graphene. Preprint: arXiv:1712.10303.

\bibitem{brehem1}
A.~Mkrtchyan, V.~Parazian, A.~Saharian, Coherent Bremsstrahlung on a deformed
  graphene sheet, Journal of Physics Conference Series 732~(012026) (2016)
  1--6.

\bibitem{brehem2}
V.~Astapenko, Y.~Krotov, Brehemshtrahlung of fast electron in graphene, Journal
  of Physics Conference Series 357~(012017) (2012) 1--8.

\bibitem{brehem3}
S.~Ktitorov, M.~RI, Electromagnetic radiation by electrons in the corrugated
  graphene, Nanosystems: Physics, Chemistry, Mathematics 7 (2012) 51--57.

\bibitem{peskin}
M.~Peskin, D.~Schroeder, An Introduction to Quantum Field Theory, Perseus Book
  Publishing, LLC, 1995.

\bibitem{hermosa} N. Hermosa, Reflection beamshifts of visible light due to
graphene, Journal of Optics {\bf 18}, (2016) 025612. 

\end{thebibliography}

\end{document}